\documentclass[12pt]{iopart}

\usepackage{iopams}
\usepackage{amssymb}
\usepackage{color}
\usepackage{graphicx}
\usepackage{amsthm,latexsym}
\newtheorem{theorem}{Theorem}

\newtheorem{lemma}[theorem]{Lemma}
\usepackage{graphicx}
\usepackage{cancel}
\usepackage{enumerate}
\usepackage{float}
\usepackage{ae}
\usepackage{aecompl}
\usepackage{enumerate}
\usepackage{soul}
\begin{document}

\title[On perturbative constraints for vacuum $f(R)$ gravity]{On perturbative constraints for vacuum $f(R)$ gravity}

\author{Daniel Molano$^{1,2}$, Fabi\'an Dar\'io Villalba$^{1,2}$\footnote{On leave from Universidad de los Andes.}, Leonardo Casta\~neda$^1$, Pedro Bargue\~no$^3$}

\address{$^1$Observatorio Astron\'omico Nacional, Universidad Nacional de Colombia, Ciudad Universitaria, Bogot\'a, Colombia\\
$^2$Departamento de F\'isica, Universidad de los Andes, CP 111711, Bogot\'a, Colombia\\
$^3$ Departamento de F\'isica Aplicada, Universidad de Alicante, Campus de San Vicente del Raspeig, E-03690 Alicante, Spain}
\ead{damolanom@unal.edu.co, fdvillalbap@bt.unal.edu.co, lcastanedac@unal.edu.co, pedro.bargueno@ua.es}


\vspace{10pt}
\begin{indented}
\item[]February 2020
\end{indented}

\begin{abstract}
Perturbative techniques are important for modified theories of gravity since they allow to calculate deviations from General Relativity without recurring to exact solutions, which can be difficult to find. When applied to models such as $f(R)$ gravity, these techniques introduce corrections in the field equations that involve higher order derivatives. Such corrections must be handled carefully to have a well defined perturbative scheme, and this can be achieved through the method of perturbative constraints, where the coefficient of the additional term in the action is used as expansion parameter for the quantities of interest. In this work, we implement a perturbative framework that compares solutions in modified theories of gravity with solutions of the Einstein field equations, by following the guidelines of perturbation theory constructed in General Relativity together with the perturbative constraints rationale. By using this formalism, we demonstrate that a consistent $f(R)$ perturbation theory in vacuum, for an important class of $f(R)$ functions, produces no additional effects with respect to what is expected from the perturbation theory of General Relativity. From this result, we argue that there are fundamental limitations that explain why the solutions of some $f(R)$ models can be disconnected from their general relativistic counterparts, in the sense that the limit that leads from the $f(R)$ action to General Relativity does not transform the solutions accordingly.
\end{abstract}

%
%
%
%
%
\section{Introduction}

\noindent The possibility that General Relativity (GR) is not the ultimate theory of gravitational phenomena has been considered from a variety of theoretical and observational arguments. Modifications of GR such as scalar-tensor, tensor-vector-scalar, and higher-order derivative gravity theories, offer an explanation for different facts that have been regarded as shortcomings of GR; for example, the  cosmological accelerated expansion could be explained with higher-order theories, without resorting to exotic sources. Among the different alternatives, $f(R)$ modified gravity (see \cite{sotiriou2010f,nojiri2011unified,nojiri2017modified} for reviews) stands out in this context for a number of theoretical properties such as the absence of instabilities and the relative freedom to describe different behaviors in cosmology and astrophysics \cite{nojiri2011unified,woodard2007avoiding}.

It is important to search for exact solutions in modified theories of gravity such as $f(R)$ models. For example, spherically symmetric solutions are very interesting since they can be compared with the well-known Schwarzschild one, and the resulting differences can be constrained from observations. However, the construction of solutions is a hard task and only some exact (and vacuum) ones are known in $f(R)$ gravity (some examples can be found in \cite{nojiri2017modified,PhysRevD.72.103005,Capozziello_2008,Sebastiani_2011}). In this context, perturbation theory approaches become important since they allow to find more realistic solutions starting from the known, exact ones. Different perturbative approaches have been developed for $f(R)$ theories of gravity, and they follow closely the approaches devised for GR given that they are relativistic theories as well. A very useful framework to obtain and understand solutions in this context was introduced by Capozziello, Stabile and Troisi for the spherically symmetric case \cite{Capozziello_2008}, in which the metric is split between a background solution plus a perturbation, and the corresponding equations for the dynamics of these parts are obtained from the $f(R)$ field equations, in a similar way than the GR case. Although this approach permits to obtain solutions in a relatively straightforward way, issues like the gauge invariance of the results and the definiteness of the perturbative scheme are not easy to consider. In fact, for any perturbative scheme, the presence of derivatives of order higher than two in the field equations of $f(R)$ gravity leads to a number of theoretical consequences, the most important of all being the presence of additional degrees of freedom. These degrees of freedom imply that the corrections introduced in the perturbative framework are not necessarily small; in addition, there is no hierarchy among the different orders of approximation, so that it could happen that the neglected terms are of the same size of the considered corrections or even larger (see \cite{simon1990higher} and references therein).

In this context, the approach of perturbative constraints (also known as order reduction), initially proposed in \cite{jaen1986reduction,eliezer1989problem} and further developed in \cite{simon1990higher,simon1991stability}, is a relevant alternative for the analysis of higher derivative theories such as $f(R)$ models. Broadly speaking, this method considers additional, higher derivative terms in the action as perturbations, and defines the perturbative expansions in terms of the constant coefficients of such terms; with these coefficients, a hierarchy of corrections is guaranteed, so that higher orders of perturbation can be safely discarded knowing that they are small. This construction leads to results such that the additional degrees of freedom present in higher derivative theories, such as $f(R)$ models, are absent in a perturbative sense. Although the formalism of perturbative constraints has been used mainly in contexts such as non-local and effective field theories (for recent examples of applications, see \cite{glavan2018perturbative,mottola2017scalar}), it also has been considered in cosmological \cite{simon1992no,dedeo2008stable,cooney2009gravity,castellanos2018higher,Solomon_2018,barros2019bouncing} and astrophysical settings \cite{cooney2010neutron}, leading to interesting insights such as a description of the accelerated cosmological expansion with no additional degrees of freedom that could be associated to dark energy. In spite of the fact that the perturbative constraints method provides a well-defined perturbative scheme to consider theories such as $f(R)$ models, the issue of gauge invariant quantities is still unsettled. The present work provides a foundation to address this issue by establishing a formalism able to undertake these questions. In this paper we study the implications of such formalism for the vacuum case, which is convenient for reasons detailed below. The important problem of the construction of a general perturbation theory in terms of gauge invariant quantities is considered in an upcoming paper \cite{molano2020}, although some relevant points regarding this matter are briefly discussed throughout the paper.

Specifically, in this work, following Bruni and coworkers \cite{bruni:1997}, we implement an approach where Modified Theories of Gravity (MTG) and GR can be compared in a similar way than general relativistic perturbation theory, with results that are suitable to compare GR (or another MTG) with Newtonian gravity through approaches such as the Post-Newtonian approximation. Our goal is to obtain a powerful way of comparing vacuum field equations and solutions for $f(R)$ MTG with their GR counterparts, hoping that such formalism can be used, in a future work \cite{molano2020}, to properly study issues such as the gauge problem in perturbation theory for this MTG. In this paper, we restrict this framework to models of the form  $f(R)=R+\lambda \Psi(R)$, where $\Psi$ is an analytic smooth function. Within this context, it is proven what we consider to be our main result: namely, that $f(R)$ vacuum perturbation theory, defined following the principles of the perturbative constraints framework for a large class of $f(R)$ functions, does not introduce additional effects with respect to the results of perturbation theory in General Relativity. From this proof, it is argued that perturbative solutions do not capture some additional features introduced by $f(R)$ gravity, so different methods or conditions beyond our theorem are to be implemented to reach the full set of $f(R)$ solutions from a general relativistic one. We must note that our results do not require any particular symmetry, so they can be of interest for current important applications of $f(R)$ models such as gravitational waves.

Our work is organized as follows: in section 2 we give a brief review of $f(R)$ gravity to state our notation and conventions. Afterwards, in section 3 we derive and discuss some known results regarding the cases with constant scalar curvature and with spherical symmetry to further motivate the results below. In section 4, we discuss the mathematical rationale behind the perturbation theory that we consider in this work; some examples in this context are examined in section 5. In this section we also state and prove the theorems relating vacuum perturbation theories for $f(R)$ MTG and GR. Finally, in section 6 we discuss our results in the context of previous works and outline some of their consequences. In section 7 we give some concluding remarks.

\section{$f(R)$ Modified Theories of Gravity}

In what follows we assume, in the same way that GR, that space-time is described by the pair $(M,g_{\mu\nu})$, where $M$ is a Hausdorff and paracompact four-dimensional manifold and $g_{\mu\nu}$ is a Lorentzian metric with signature $-+++$ on $M$. The field equations for $g_{\mu\nu}$ can be obtained from the variation of the action:
\begin{equation}\label{eq:accionfr}
S(g)=\int_\mathcal{V} \mathcal{L}\, dx^4 +S_B,
\end{equation}
where
\begin{equation*}
\mathcal{L}=\frac{1}{16\pi G}f(R) \sqrt{-g}+\mathcal{L}_M(g_{\mu\nu},\varphi,\varphi_{,\mu})\sqrt{-g}, \end{equation*}
accounts for a generalization of the Einstein-Hilbert term through a general function $f(R)$ of the Ricci scalar, whereas the second term corresponds to the matter fields ($c=1$). $S_B$ is the corresponding boundary term (see \cite{raey} for more details). In the context of the (metric) variational principle for this $f(R)$ MTG action, $g_{\mu\nu}$ is a solution of the field equations,
\begin{equation}
 f'(R)R_{\mu\nu}-\frac{1}{2}f(R)g_{\mu\nu}-\nabla_{\mu} \nabla_{\nu} f'(R)
+g_{\mu\nu}\Box f'(R)=8\pi G T_{\mu\nu},
 \label{fieldeq}
\end{equation}
where $f'(R)\equiv \partial f/ \partial R$, $T_{\mu\nu}$ is the energy-momentum tensor that describes the matter distribution,  $R \equiv R_{\mu\nu}g^{\mu\nu}$ is the curvature scalar and the Ricci tensor is $R_{\mu\nu}\equiv {R_{\mu\rho\nu}}^\rho$. The Riemann tensor in a local coordinate basis is given by
\begin{equation}{R_{\mu\nu\rho}}^\sigma=\frac{\partial}{\partial x^\nu}{\Gamma^\sigma}_{\mu\rho}-\frac{\partial}{\partial x^\mu}{\Gamma^\sigma}_{\nu\rho}+{\Gamma^\alpha}_{\mu\rho}{\Gamma^\sigma}_{\alpha\nu}-{\Gamma^\alpha}_{\nu\rho}{\Gamma^\sigma}_{\alpha\mu},
\end{equation}
where ${\Gamma^\sigma}_{\mu\rho}$ are the Christoffel symbols which can be derived from the metric tensor through the expression:
$${\Gamma^\rho}_{\mu\nu}=\frac{1}{2}g^{\rho\sigma}\biggl[\frac{\partial g_{\nu\sigma}}{\partial x^\mu}+\frac{\partial g_{\mu\sigma}}{\partial x^\nu}-\frac{\partial g_{\mu\nu}}{\partial x^\sigma}\biggr].$$
Finally, $f:\mathbb{R}\longrightarrow\mathbb{R}$ is a free function which we consider to be $C^\infty$ and analytic. These constraints could be weaker but we need them in order to have well-defined Taylor expansions. Notably, if $f(R)=R$, we recover Einstein's gravity without a cosmological constant.

We denote the geometrical sector of the Einstein field equations with $G_{\mu\nu}$, and the tensor that represents the geometrical sector for the equation (\ref{fieldeq}) with $\Sigma_{\mu\nu}$. That is,
\begin{equation}
\Sigma_{\mu\nu}\equiv f'(R)R_{\mu\nu}-\frac{1}{2}f(R)g_{\mu\nu}-\nabla_{\mu} \nabla_{\nu} f'(R)+g_{\mu\nu}\Box f'(R).
\end{equation}
It is worth-noticing that the $f(R)$ field equations (\ref{fieldeq}) are equivalent to a set of non-linear fourth-order partial differential equations on the metric components $g_{\mu\nu}$. Therefore, they require a boundary specification of the metric and their first three derivatives to determine a solution. This requirement evidently contrasts with the GR case, where only boundary data for the metric and its first derivatives are required. This fact will have important consequences when considering GR as a limit of $f(R)$ MTG, as discussed below.

\section{Some results for vacuum and spherically symmetric field equations}

In this section, we show some analytical results concerning vacuum and spherically symmetric field equations, which allow us to obtain analogous expressions for results widely known in GR. Firstly, we assume a space time with a constant and uniform Ricci curvature $R=R_0$.

\begin{theorem}
Let $f:\mathbb{R} \longrightarrow \mathbb{R}$ be a function such that $f\in \mathcal{C}^r$, $r\geq 1$, and  $f'(R)\neq0$. Also, suppose a constant Ricci scalar with constant value $R=R_0$. Then, the field equations for $f(R)$ MTG in vacuum are reduced to two possibilities:
\begin{enumerate}
\item GR field equations without cosmological constant if $R_0=0$.
\item GR field equations with non-vanishing cosmological constant if $R_0\neq0$.
\end{enumerate}
\end{theorem}

Proof: If we have a constant and uniform $R=R_0$, then $f(R_0)$ and $f'(R_0)$ are also constant and uniform, thus  $\nabla_\mu\nabla_\nu f'(R)=0$, $\Box f'(R)=0$. In this way, the field equations and their corresponding trace are, in vacuum
\begin{equation}\label{fieldvacio}
 f'(R_0)R_{\mu\nu}-\frac{1}{2}f(R_0)g_{\mu\nu}=0,
\end{equation}
\begin{equation}\label{trazavacio}
f'(R_0)R_0-2f(R_0)=0.
\end{equation}
\begin{enumerate}
\item Then, if $R_0=0$, $f(R_0)=0$ because of (\ref{trazavacio}), replacing this equation in (\ref{fieldvacio}) we have $R_{\mu\nu}=0$, which corresponds to GR without cosmological constant.
\item If $R_0\neq0$, then $f(R_0)=f'(R_0)R_0/2$ by (\ref{trazavacio}). Thus, given (\ref{fieldvacio}), we have
\begin{eqnarray*}
 f'(R_0)R_{\mu\nu}-\frac{1}{2}\left(\frac{f'(R_0)R_0}{2}\right)g_{\mu\nu}=0,\\
f'(R_0)\left(R_{\mu\nu}-\frac{1}{4}R_0 g_{\mu\nu}\right)=0,
\end{eqnarray*}
or, considering that $f'(R)\neq0$,
\begin{equation}
R_{\mu\nu}-\frac{1}{4}R_0 g_{\mu\nu}=0,
\end{equation}
which corresponds to the GR field equations with cosmological constant $\Lambda=\frac{1}{4}R_0$. $\blacksquare$.
\end{enumerate}

If the case $f'(R)=0$ is considered, then the equation (\ref{trazavacio}) implies that $f(R)$ is also zero and, by equation (\ref{fieldvacio}), no field equation exists. The only information of the system is present in the Ricci scalar. In this case, the variational principle would be an ill-defined approach to obtain field equations. 

It is interesting to consider this result from the point of view of boundary conditions for the differential equation (\ref{fieldeq}). As we mentioned before, this system consists of fourth-order differential equations, hence, its solution requires the specification of the second- and third- order derivatives of the metric on the boundary. This requirement is equivalent to demand boundary conditions for the Ricci scalar and its first derivative. Therefore, the constant value $R_0$ that we considered in Theorem 1 must comply with the respective boundary conditions, and in this way, we can interpret the cosmological constant as a parameter determined by boundary conditions, as some authors have suggested \cite{padmanabhan2016atoms}.

Now, let us consider the static, spherically symmetric case. To do this, it is useful to develop a new version of the field equations as follows. The trace of (\ref{fieldeq}) leads to the following expression for the D'Alembertian of $f'(R)$:

\begin{equation}\label{trace}
 \square f'(R)=\frac{8\pi G T+2f(R)-f'(R)R}{3},
\end{equation}
where $T$ is the trace of the energy-momentum tensor. Replacing this result in (\ref{fieldeq}) and simplifying, a new version for the $f(R)$ field equations (\ref{fieldeq}) is obtained:
\begin{equation}\label{fieldeq2a}
 f'R_{\mu\nu}+\frac{g_{\mu\nu}}{3}\left(\frac{1}{2}f-f'R+8\pi G T\right)-\nabla_\mu\nabla_\nu f'=8\pi G T_{\mu\nu}.
\end{equation}
It must be noted that this equation leads to the GR field equations when $f(R)=R$. Explicitly, replacing this form for $f(R)$ (and  $f'(R)=1$, a constant) in (\ref{fieldeq2a}) we obtain:
\begin{equation}\label{grfieldprime}
R_{\mu\nu}+\frac13 g_{\mu\nu}\left(8\pi G T-\frac12 R\right)=8\pi G T_{\mu\nu}.
\end{equation}
This equation is equivalent to Einstein field equations, which can be shown as follows: the trace of (\ref{grfieldprime}) leads to the relation
\begin{equation}
8\pi G T=-R,
\end{equation}
which already agrees with GR; furthermore, replacement of this result for $T$ in (\ref{grfieldprime}) leads, after a simplification, to the Einstein field equations,
\begin{equation}
R_{\mu\nu}+\frac12 g_{\mu\nu}R=8\pi G T_{\mu\nu}.
\end{equation}
In what follows, we focus on the vacuum version of (\ref{fieldeq2a}), namely:
\begin{equation}\label{fieldeq2}
 f'R_{\mu\nu}+\frac{g_{\mu\nu}}{3}\left(\frac{1}{2}f-f'R\right)-\nabla_\mu\nabla_\nu f'=0.
\end{equation}

A general static and spherically symmetric metric for a pseudo-Riemannian manifold, where the radial coordinate $r$ is defined as the areal radius and the coordinates are chosen to obtain a diagonal metric, can be written as \cite{wald:1984}:
\begin{equation}
 ds^2=-e^{\eta(r)} dt^2+ e^{\alpha(r)} dr^2+r^2d\Omega^2.
\end{equation}
Now, by inserting this expression in (\ref{fieldeq2}), we obtain,
\begin{eqnarray*}
 0=\Sigma_{00}=&e^\eta\biggl\{\biggl[\frac{e^{-\alpha}}{4}(2\ddot{\eta}+\dot{\eta}^2-\dot{\eta}\dot{\alpha})+\frac{e^{-\alpha}\dot{\eta}}{r}\biggr]f'\\
&-\frac{1}{3}\biggl(\frac{1}{2}f-f'R\biggr)+\frac{1}{2}\dot{\eta}e^{-\alpha}\dot{R}f''\biggr\},\\
0=\Sigma_{11}=&\biggl[-\frac{1}{4}(2\ddot{\eta}+\dot{\eta}^2-\dot{\eta}\dot{\alpha})+\frac{\dot{\alpha}}{r}\biggr]f'\\
&+\frac{e^{\alpha}}{3}\biggl(\frac{1}{2}f-f'R\biggr)-f'''\dot{R}^2-f''\ddot{R}+\frac{1}{2}\dot{\alpha}\dot{R}f'',\\
0=\Sigma_{22}=&\biggl(1+\frac{-\dot{\eta}r+\dot{\alpha}r-2}{2}e^{-\alpha}\biggr)f'\\
&+\frac{r^2}{3}\biggl(\frac{1}{2}f-f'R\biggr)-r\dot{R}f''e^{-\alpha},
\end{eqnarray*}
where $\dot{}\equiv\frac{d}{dr}$. If we consider the following combination of equations,
\begin{eqnarray*}
 0=e^{-\eta}\Sigma_{00}+e^{-\alpha}\Sigma_{11}=&e^{-\alpha}\biggl[\frac{f'}{r}(\dot{\eta}+\dot{\alpha})+\frac{1}{2}(\dot{\eta}+\dot{\alpha})f''\dot{R}\\
&-f'''\dot{R}^2-f''\ddot{R}\biggr]
\end{eqnarray*}
we obtain,
\begin{equation}\label{ec:70}
 \dot{\eta}+\dot{\alpha}=\frac{f'''\dot{R}^2+f''\ddot{R}}{\frac{f'}{r}+\frac{f''\dot{R}}{2}}=\frac{2r\partial_r^2f'}{2f'+r\partial_r f'}.
\end{equation}
The right-hand side of the last equation is zero for GR, so that it reduces, in this case, to the standard equation that leads to the Schwarzschild solution. Although some authors (for example in \cite{Sebastiani2011}) propose particular $f(R)$ models to cancel out this factor of Eq. (\ref{ec:70}) in order to obtain exact solutions within spherical symmetry, it is quite difficult to obtain this kind of solutions for relevant general proposals.

Given this situation, a variety of perturbative schemes has been devised in order to obtain corrections to exact solutions such as the Schwarzschild metric. Some of these schemes, as discussed above, are based on considering $f(R)$ functions that can be written as the Einstein-Hilbert function $R$ plus a correction which is small because of a coefficient $\lambda\ll1$, i.e. $f(R)=R+\lambda \Psi(R)$. One then considers that all quantities of interest can be expanded as power series on $\lambda$, whose truncation leads to the different orders of approximation. This constitutes the basis of the so-called perturbative constraints approach or order reduction \cite{jaen1986reduction,eliezer1989problem,simon1990higher}. Our work is framed in this line of development, as we show in the following sections; however, it is relevant for our future discussion to mention another possibility, that the metric elements themselves can be considered as a sum of a background solution plus a quantity regarded as a small perturbation, so that $g_{\mu\nu}=g^{(backg)}_{\mu\nu}+g^{(pert)}_{\mu\nu}$. This is the approach followed by Capozziello, Stabile, and Troisi \cite{Capozziello_2008}. With these possibilities in mind, we construct a general scheme, following \cite{bruni:1997}, to implement a perturbation theory for $f(R)$ theories in a precise mathematical form in the following section.

\section{$f(R)$ Perturbation Theory}

Most theories in nature can be modeled with a system of ordinary or partial differential equations; however, we do not have a general method to obtain general solutions for all of them. As we illustrated above, for $f(R)$ MTG, when the function $f$ is not linear we have field equations in terms of non-linear fourth-order partial differential equations, whose exact solutions are known only for some specific functions $f(R)$ and special symmetric conditions \cite{zbMATH05782036}. Furthermore, in some cases of interest in which we do not know the exact solution in $f(R)$ MTG, it is not possible to compare with GR nor suppose that the solutions are 'close' to each other if $f(R)$ is 'close' to $R$. For example, let us suppose that we have two theories like $f_0(R)=R$ and $f_\lambda(R)=R+\lambda R^2$, which correspond to GR and the Starobinsky model \cite{Starobinsky198099}, respectively. If $\lambda=0$ we recover GR, and it can be expected that if $g_\lambda$ is solution of the model $f_\lambda$, this solution tends to a GR solution when $\lambda$ tends to zero. Nonetheless,  this assumption is not always correct as shown by explicit examples \cite{cikintouglu2018vacuum}.

\
We appeal to the essence of perturbation theory in GR to construct a framework that allows for comparisons between solutions in $f(R)$ MTG and GR. Let us consider a 4+1 dimensional manifold $\mathcal{N}$ which is foliated by space-time manifolds $M_\lambda$. Therefore, $\mathcal{N}=M\times\mathbb{R}$. For convenience, we denote $M_\lambda\equiv M\times \{\lambda\}$. On each slice $M_\lambda$ we have a metric $g_\lambda$ which is a solution of some model of gravity, whereas in $M_0$ we have a metric $g_0$ which is a GR solution. In this context, we suppose that $\mathcal{N}$ is well defined and constitutes a cobordism between the manifolds $M_0$ and $M_\lambda$. The necessity of this construction will be explained in short. We suppose that different models of gravity are linked continuously by the value of $\lambda$. As an example of the latter in $f(R)$ MTG we can point out the Starobinsky model, $f(R)=R+\lambda R^2$.

\begin{figure}[H]
\begin{center}
\includegraphics[width=0.8\linewidth]{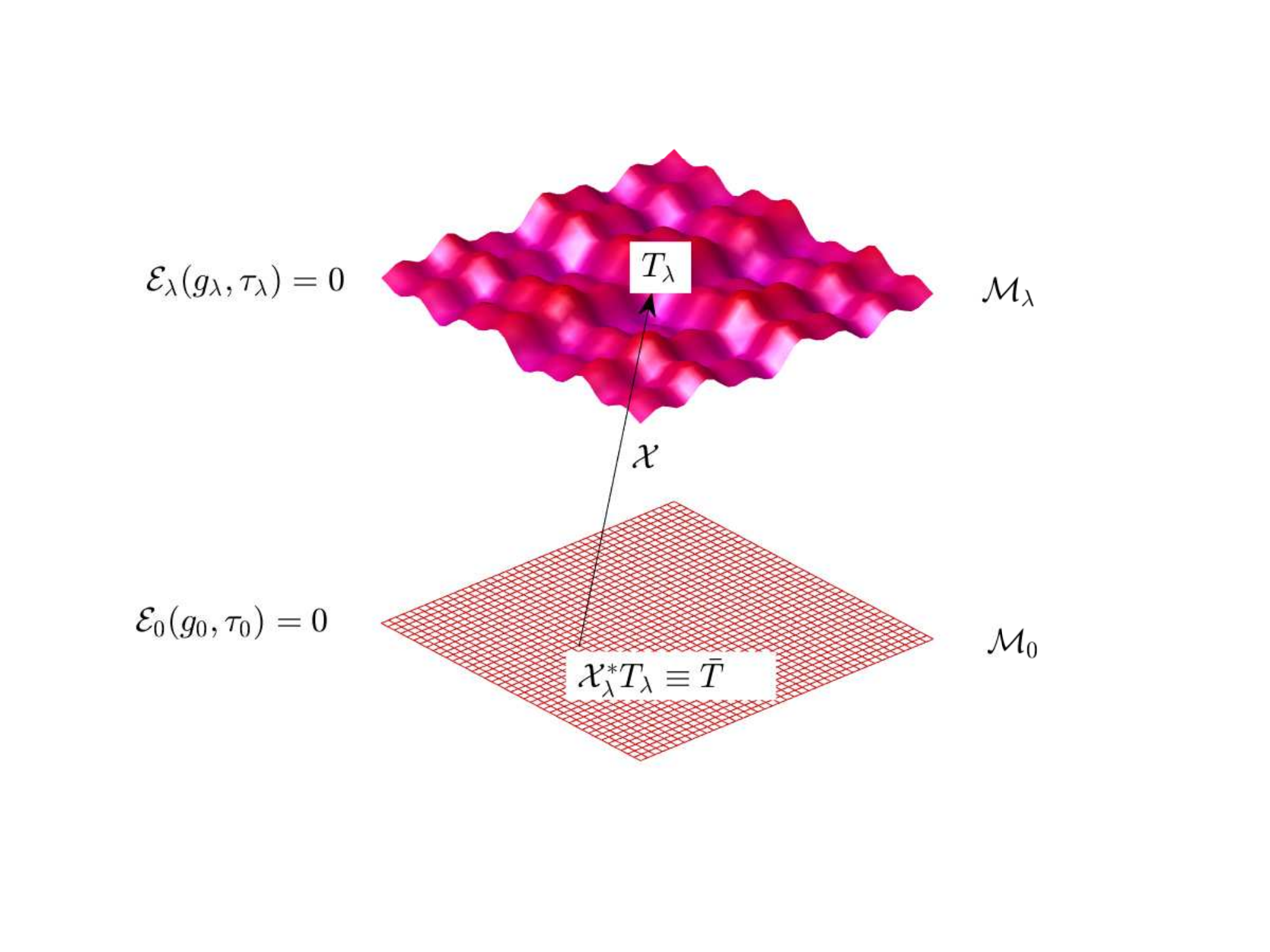}
\end{center}
\caption{The (4+1)-dimensional manifold $\mathcal{N}$. On each $M_\lambda$ we have a model of gravity, whereas the model in $M_0$ is GR. In perturbation theory, $M_0$ is called \emph{background space-time.}}
\label{fig:variedadntgmpe}
\end{figure}

In more precise terms \cite{wald:1984}, we consider a generic field equation
\begin{equation*}
 \mathcal{E}_0(g_0,\tau_0)=0,
\end{equation*}
here, $g_0$ is a exact solution of $\mathcal{E}_0$ together with a stress-energy source $\tau_0$, and constitutes the so-called background. Now, we suppose that we can build a one-parametric family, continuous, differentiable and analytic of exact solutions $g_\lambda$, such that
\begin{equation}
 \mathcal{E}_\lambda(g_\lambda,\tau_\lambda)=0
\end{equation}
where $\mathcal{E}_\lambda$ are the field equations for the perturbed theory, for example MTG, and $g_\lambda$ is the metric solution for such model. It must be noted that we allowed a dependence of the stress-energy tensor $\tau$ on $\lambda$ by writing $\tau_\lambda$, which is motivated mainly by two considerations: firstly, this tensor depends implicitly on this parameter through the metric $g_\lambda$, and, more importantly, there are models where the corrections to the field equations can be regarded as effective corrections to the stress energy tensor \cite{zbMATH05782036}.  Figure \ref{fig:variedadntgmpe} summarizes this geometrical construction.

In the context of $f(R)$ MTG, it is important to note, recalling the discussion of the previous section, that the class of field equations described by $\mathcal{E}_\lambda$ is to be supplemented by a set of boundary conditions, which are to be imposed even for $\lambda\rightarrow0$. The unperturbed solution might not satisfy such boundary conditions since the corresponding equations do not involve the higher order derivatives. This issue is known as the presence of boundary layers which condition the possible solutions for $\lambda\rightarrow0$. In fact, as illustrated by \cite{cikintouglu2018vacuum}, the resulting $g_0$ is not a solution of Einstein's equations (which are the $\lambda\rightarrow0$ limit of Starobinsky model's field equations), but it is a metric with pathological features such as non-differentiability and a discontinuity, all of them induced by the boundary layer present in the problem. This issue reflects the problems involved when obtaining limits of spacetimes \cite{geroch:1969} and needs to be considered when constructing a perturbation theory around the background solution $g_0$.

Following the structure of GR perturbation theory, we suppose a smooth and analytic tensor field $T$ on $\mathcal{N}$ and we want to compare tensor fields $T_\lambda$ on $M_\lambda$ with $T_0$ on $M_0$ through $\mathcal{N}$. With such objective, we are going to map elements on $M_\lambda$ with elements on $M_0$, this mapping is implemented with a \emph{flow} $\mathcal{X}_\lambda:\mathcal{N}\rightarrow\mathcal{N}$, such that it is generated by a vector field $X$, which is transversal to each submanifold $M_\lambda$ and rotationless. In a specific chart we can choose the coordinate $X^5$ to be a constant as long as it is different from zero (a zero value would imply that the flux is restricted to $M_0$); nevertheless, it must be remarked that $X$ can be arbitrary except for the requirements aforementioned. Within this construction, we can expand the pull-back $\mathcal{X}_\lambda^* T$ of the tensor field $T$ around $\lambda=0$ with a Taylor expansion, which is defined in terms of the Lie derivatives along $\mathcal{X}$ of the tensor fields of interest (see \cite{bruni:1997,nakamura2019second} for additional details regarding this expansion):
\begin{equation}\label{eq:taylorexpansion}
\mathcal{X}_\lambda^* T=\sum_{k=0}^\infty \frac{\lambda^k}{k!}\pounds_X^k T,
\end{equation}
where $\pounds_X^k T$ denotes the $k$-th Lie derivative of the tensor field $T$ along the flow generated by $X$. Assuming that any tensor field has an expansion of the form (\ref{eq:taylorexpansion}), we provide the basis for the perturbative constraints method, since the truncation of the series at some order $n$ provides approximations for the quantities of interest of the theory. It must be remarked that the expansion parameter is the same one that multiplies the correction to the Einstein-Hilbert action. This fact has profound consequences in the framework of perturbative constraints \cite{simon1990higher} that will manifest in our results, such as the absence of the new degrees of freedom that might produce boundary layers. More details regarding this feature are discussed below.

If we denote the pull-back $\mathcal{X}_\lambda^* T$ with $\bar{T}$ and use the abbreviation $\mathop{T}\limits^{(k)}\equiv \pounds_X^k T$ for the Lie derivatives, then we have
\begin{equation}\label{eq:tensorbar3}
 \bar{T}=\mathop{T}\limits^{(0)}+\lambda\mathop{T}\limits^{(1)}+\frac{\lambda^2}{2!}\mathop{T}\limits^{(2)}+\cdots.
\end{equation}

At this point, with the geometric tools introduced, we can further discuss the necessity of the manifold $\mathcal{N}$. Recalling previous work in perturbation theory by Bruni and coworkers \cite{bruni:1997}, reviewed recently in \cite{nakamura2019second}, the usual construction involves flows which are defined within the four-dimensional manifold $M$. These flows are thus automorphisms on this manifold and, by general covariance, must keep invariant the field equations of GR. However, since we are comparing a $f(R)$ theory with GR and they have different field equations, the flow $X$ that links them must be a diffeomorphism which acts on a dimension apart from the usual space dimensions. From this analysis, general covariance can be stated in this context as diffeomorphism invariance {\it within} each slice $M_\lambda$, and we expect that transformations involving the additional direction are outside the range of validity of this principle.

Another important point that can be considered within the implemented perturbation theory is the gauge freedom of the second kind \cite{sachs:1964}. It can be summarized as follows: we have different ways to identify points in $M_\lambda$ with the background; in other words, there are infinite gauge choices $X$. When we bring tensor quantities to the background from different gauges, they do not always remain invariant. One possible condition to guarantee the invariance is that the background quantities vanish, this fact is a consequence of the generalized Stewart and Walker lemma (for details, see \cite{bruni:1997}) and will be very important to establish the gauge invariance of our results, as shown below. Let us remark that the notation introduced in Eq. (\ref{eq:tensorbar3}) does not take into account explicitly the flux $\mathcal{X}$ nor the vector field $X$, and this anticipates the gauge invariance aforementioned.

\section{Relation between $f(R)$ MTG and GR perturbation theories}
In this section we state and demonstrate the main result of this work: the equivalence of $f(R)$ and GR perturbation theories in absence of matter for a specific class of $f(R)$ models.

Let $\bar{P}$ and $\bar{Q}$ be two tensor fields such that they have an associated expansion of the form (\ref{eq:tensorbar3}), that is,
\begin{eqnarray}
 \bar{P}&=\mathop{P}\limits^{(0)}+\lambda\mathop{P}\limits^{(1)}+\frac{\lambda^2}{2!}\mathop{P}\limits^{(2)}+\cdots,\\
 \bar{Q}&=\mathop{Q}\limits^{(0)}+\lambda\mathop{Q}\limits^{(1)}+\frac{\lambda^2}{2!}\mathop{Q}\limits^{(2)}+\cdots.
\end{eqnarray}
Immediately, it can be shown that,
\begin{equation}
\bar{P}\bar{Q}=\sum_{n=0}^\infty \frac{\lambda^n}{n!}\sum_{i=0}^n{n \choose i} \mathop{P}\limits^{(i)}\mathop{Q}\limits^{(n-i)}.
\end{equation}
Thus, the $n$-th term of the tensor product $\bar{P}\bar{Q}$ is written as
\begin{equation}\label{expmult}
 \mathop{PQ}\limits^{(n)}\equiv\sum_{i=0}^n{n \choose i}\mathop{P}\limits^{(i)}\mathop{Q}\limits^{(n-i)},
\end{equation}
where
\begin{equation}
 {n \choose i}=\frac{n!}{i!(n-i)!}.
\end{equation}
Using this result in the geometrical sector of $f(R)$ MTG field equations:
\begin{equation}
 \bar{\Sigma}_{\mu\nu}=\bar{f'}\bar{R}_{\mu\nu}-\frac{1}{2}\bar{f}\bar{g}_{\mu\nu}-\bar{\nabla}_\mu\bar{\nabla}_\nu\bar{f'}+\bar{g}_{\mu\nu}\bar{\Box}\bar{f'},
\end{equation}
we introduce the following expansions,
\begin{eqnarray}
 \bar{f}&=\mathop{f}\limits^{(0)}+\lambda\mathop{f}\limits^{(1)}+\frac{\lambda^2}{2!}\mathop{f}\limits^{(2)}+\cdots\\
\bar{f'}&=\mathop{f'}\limits^{(0)}+\lambda\mathop{f'}\limits^{(1)}+\frac{\lambda^2}{2!}\mathop{f'}\limits^{(2)}+\cdots\\\label{expmetrica}
\bar{g}_{\mu\nu}&=\mathop{g_{\mu\nu}}\limits^{(0)}+\lambda\mathop{g_{\mu\nu}}\limits^{(1)}+\frac{\lambda^2}{2!}\mathop{g_{\mu\nu}}\limits^{(2)}+\cdots\\
\bar{R}_{\mu\nu}&=\mathop{R_{\mu\nu}}\limits^{(0)}+\lambda\mathop{R_{\mu\nu}}\limits^{(1)}+\frac{\lambda^2}{2!}\mathop{R_{\mu\nu}}\limits^{(2)}+\cdots.
\end{eqnarray}
and the identity (see \cite{wald:1984})
\begin{equation}
 \bar{\nabla}_\mu\bar{\nabla}_\nu\bar{f'}=\nabla_\mu\nabla_\nu \bar{f'}-{C_{\mu\nu}}^\delta\nabla_\delta \bar{f'},
\end{equation}
where
\begin{equation}
 {C_{\mu\nu}}^\delta=\frac{1}{2}\bar{g}^{\alpha\delta}(\nabla_\mu\bar{g}_{\alpha\nu}+\nabla_\nu\bar{g}_{\alpha\mu}-\nabla_\alpha\bar{g}_{\mu\nu}).
\end{equation}
Together with the property (\ref{expmetrica}), we can make the following expansion:
\begin{equation}
 {C_{\mu\nu}}^\delta =\mathop{{C_{\mu\nu}}^\delta}\limits^{(0)}+\lambda\mathop{ {C_{\mu\nu}}^\delta}\limits^{(1)}+\frac{\lambda^2}{2!}\mathop{ {C_{\mu\nu}}^\delta}\limits^{(2)}+\cdots
\end{equation}
We have $\nabla_{a}\mathop{g_{bc}}\limits^{(0)}=0$ in the background, then $\mathop{{C_{\mu\nu}}^\delta}\limits^{(0)}=0$. With these results, we can obtain the $n$-th term of the geometrical sector of the field equation expansion:
\begin{equation}\label{ecufrn}
 \eqalign{\mathop{\Sigma_{\mu\nu}}\limits^{(n)}=&\sum_{i=0}^n\biggl[{n \choose i}\mathop{f'}\limits^{(i)}\mathop{R_{\mu\nu}}\limits^{(n-i)}-\frac{1}{2}{n \choose i}\mathop{f}\limits^{(i)}\mathop{g_{\mu\nu}}\limits^{(n-i)}\biggr]\\
 &-\nabla_\mu\nabla_\nu \mathop{f'}\limits^{(n)}+\sum_{i=0}^n{n \choose i}\mathop{{C_{\mu\nu}}^\alpha}\limits^{(n-i)}\nabla_\alpha\mathop{f'}\limits^{(i)}\\
&+\sum_{i=0}^n\sum_{k=0}^i {n \choose i}{i \choose k}\mathop{g_{\mu\nu}}\limits^{(n-i)}\mathop{g^{\alpha\beta}}\limits^{(i-k)}\\
&\cdot\biggr[\nabla_\alpha\nabla_\beta \mathop{f'}\limits^{(k)}-\sum_{l=0}^k {k \choose l}\mathop{{C_{\alpha\beta}}^\delta}\limits^{(k-l)}\nabla_\delta\mathop{f'}\limits^{(l)}\biggr].}
\end{equation}
With this expression, we can now consider a particular $f(R)$ model and study their consequences in the vacuum case. Firstly, we explore the model $f(R)=R+\lambda R^2$, and then, we generalize to the model $f(R)=R+\lambda\Psi_\lambda(R)$. It is important to remark that we assume that these models are exact, so we are not neglecting further terms in the action that could produce inconsistencies in our perturbation theory \cite{simon1990higher,castellanos2018higher}.

\subsection{$f(R)=R+\lambda R^2$ case}
For the model $f(\bar{R})=\bar{R}+\lambda {\bar{R}}^2$, we have $f'(\bar{R})=1+2\lambda \bar{R}$, where $\bar{R}=\mathop{R}\limits^{(0)}+\lambda\mathop{R}\limits^{(1)}+\frac{\lambda^2}{2!}\mathop{R}\limits^{(2)}+\cdots$, hence

\begin{eqnarray}\label{f0}
          \mathop{f}\limits^{(0)}&=\mathop{R}\limits^{(0)}  \mbox{ for  } n=0\\\label{fn}
          \mathop{f}\limits^{(n)}&=\mathop{R}\limits^{(n)}+n\sum_{i=0}^{n-1}{n-1 \choose i}\mathop{R}\limits^{(i)}\mathop{R}\limits^{(n-i-1)}  \mbox{ for  } n\geq 1
\end{eqnarray}

and

\begin{eqnarray}\label{f'0}
          \mathop{f'}\limits^{(0)}&=1  \mbox{ for  } n=0.\\ \label{f'n}
          \mathop{f'}\limits^{(n)}&=2n\mathop{R}\limits^{(n-1)}  \mbox{ for  } n\geq 1.
\end{eqnarray}
Let us remark, to avoid confusion regarding the derivatives, that Eqs. (\ref{f'0}) and (\ref{f'n}) correspond to the Lie derivatives of the function $f'(R)$, which is constructed by derivating $f(R)$ with respect to its argument, as discussed above.
\begin{lemma}\label{lema1}
Let $\mathop{R_{\mu\nu}}\limits^{(i)}=0$ for all $i=1,\dots,n-1$. Then
$\mathop{R}\limits^{(n)}=\mathop{R_{\alpha\beta}}\limits^{(n)}\mathop{g^{\alpha\beta}}\limits^{(0)}$
\end{lemma}
Proof: As we know, $\bar{R}=\bar{R}_{\alpha\beta}\bar{g}^{\alpha\beta}$
 and by equation (\ref{expmult}) we have

\begin{equation}
 \mathop{R}\limits^{(n)}=\sum_{i=0}^{n} {n \choose i} \mathop{R_{\alpha\beta}}\limits^{(i)}\mathop{g^{\alpha\beta}}\limits^{(n-i)},
\end{equation}
but, if $\mathop{R_{\alpha\beta}}\limits^{(i)}=0$ for all $i=1,\dots,n-1$, all terms vanish except one, and we obtain the desired expression
\begin{equation}
 \mathop{R}\limits^{(n)}= \mathop{R_{\alpha\beta}}\limits^{(n)}\mathop{g^{\alpha\beta}}\limits^{(0)}.
\end{equation}

\begin{theorem}
Let $\bar{\Sigma}_{ab}=0$ be the vacuum field equations for $f(R)$ MTG in the model $f(\bar{R})=\bar{R}+\lambda \bar{R}^2$, then $\bar{\Sigma}_{ab}=\bar{G}_{ab}$ in vacuum.
 \end{theorem}
Proof: We first show that $\mathop{\Sigma_{\mu\nu}}\limits^{(n)}=\mathop{G_{\mu\nu}}\limits^{(n)}$ by induction. We first consider the following cases:
\begin{enumerate}
 \item $n=0$ case,
\begin{equation}
\mathop{\Sigma_{\mu\nu}}\limits^{(0)}=\mathop{f'}\limits^{(0)}\mathop{R_{\mu\nu}}\limits^{(0)}-\frac{1}{2}\mathop{f}\limits^{(0)}\mathop{g_{\mu\nu}}\limits^{(0)}-\nabla_\mu\nabla_\nu\mathop{f'}\limits^{(0)}+\mathop{g_{\mu\nu}}\square\mathop{f'}\limits^{(0)}.
\end{equation}
Replacing  (\ref{f0}) and (\ref{f'0}), it follows that the terms involving the D'Alembertian and the double covariant derivative vanish:
\begin{eqnarray}
\mathop{\Sigma_{\mu\nu}}\limits^{(0)}&=&\mathop{R_{\mu\nu}}\limits^{(0)}-\frac{1}{2}\mathop{R}\limits^{(0)}\mathop{g_{\mu\nu}}\limits^{(0)}-\nabla_\mu\nabla_\nu (1)+\mathop{g_{\mu\nu}}\square (1)\nonumber\\
&=&\mathop{R_{\mu\nu}}\limits^{(0)}-\frac{1}{2}\mathop{R}\limits^{(0)}\mathop{g_{\mu\nu}}\limits^{(0)}=\mathop{G_{\mu\nu}}\limits^{(0)}.
\end{eqnarray}
By taking the trace of this equation, and using the vacuum assumption, we obtain $\mathop{R_{\mu\nu}}\limits^{(0)}=0$ and $\mathop{R}\limits^{(0)}=0$.
\item
\subitem $n=1$ case,

Using (\ref{ecufrn}) we have
\begin{eqnarray*}
 \mathop{\Sigma_{\mu\nu}}\limits^{(1)}=&\mathop{f'}\limits^{(1)}\mathop{R_{\mu\nu}}\limits^{(0)}+\mathop{f'}\limits^{(0)}\mathop{R_{\mu\nu}}\limits^{(1)}-\frac{1}{2}\bigl(\mathop{f}\limits^{(1)}\mathop{g_{\mu\nu}}\limits^{(0)}+\mathop{f}\limits^{(0)}\mathop{g_{\mu\nu}}\limits^{(1)})\\
 &-\nabla_\mu\nabla_\nu\mathop{f'}\limits^{(1)}+\mathop{{C_{\mu\nu}}^\alpha}\limits^{(1)}\nabla_\alpha\mathop{f'}\limits^{(0)}\\
 &+\mathop{g_{\mu\nu}}\limits^{(1)}\square\mathop{f'}\limits^{(0)}+\mathop{g_{\mu\nu}}\limits^{(0)}\mathop{g^{\alpha\beta}}\limits^{(1)}\nabla_\alpha\nabla_\beta\mathop{f'}\limits^{(0)}\\
 &+\mathop{g_{\mu\nu}}\limits^{(0)} \mathop{g^{\alpha\beta}}\limits^{(0)}[\nabla_\alpha\nabla_\beta\mathop{f'}\limits^{(1)}+\mathop{{C_{\alpha\beta}}^\delta}\limits^{(1)}\nabla_\delta\mathop{f'}\limits^{(0)}]
\end{eqnarray*}
Because of the $n=0$ results we have $\mathop{R_{\mu\nu}}\limits^{(0)}=0$ and $\mathop{R}\limits^{(0)}=0$. Replacing these results and Eqs. (\ref{fn}) and (\ref{f'n}), we obtain,
\begin{equation}
\mathop{\Sigma_{\mu\nu}}\limits^{(1)}= \mathop{R_{\mu\nu}}\limits^{(1)}-\frac{1}{2}\mathop{R}\limits^{(1)}\mathop{g_{\mu\nu}}\limits^{(0)}=\mathop{G_{\mu\nu}}\limits^{(1)}.
\end{equation}
An straightforward calculation of the trace and using Lemma \ref{lema1} imply that $\mathop{R_{\mu\nu}}\limits^{(1)}=0$ and $\mathop{R}\limits^{(1)}=0$.

\item Now by induction hypothesis we have $\mathop{\Sigma_{\mu\nu}}\limits^{(i)}=\mathop{G_{\mu\nu}}\limits^{(i)}$ for all $i=0,\dots,n$ and we want to prove that $\mathop{\Sigma_{\mu\nu}}\limits^{(n+1)}=\mathop{G_{\mu\nu}}\limits^{(n+1)}$. In addition, the trace of the equations $\mathop{G_{\mu\nu}}\limits^{(i)}=0$ shows that  $\mathop{R_{\mu\nu}}\limits^{(i)}=0$ and $\mathop{R}\limits^{(i)}=0$ for all $i=0,\dots,n$. From Eq.  (\ref{ecufrn}), isolating the terms with $\mathop{f}\limits^{(0)}$ and $\mathop{f'}\limits^{(0)}$, we have
\end{enumerate}

\begin{eqnarray*}
 \mathop{\Sigma_{\mu\nu}}\limits^{(n)}=&\sum_{i=1}^n\biggl[{n \choose i}\mathop{f'}\limits^{(i)}\mathop{R_{\mu\nu}}\limits^{(n-i)}-\frac{1}{2}{n \choose i}\mathop{f}\limits^{(i)}\mathop{g_{\mu\nu}}\limits^{(n-i)}\biggr]\\
 &-\nabla_\mu\nabla_\nu \mathop{f'}\limits^{(n)}+\sum_{i=1}^n{n \choose i}\mathop{{C_{\mu\nu}}^\alpha}\limits^{(n-i)}\nabla_\alpha\mathop{f'}\limits^{(i)}\\
&+\sum_{i=1}^n\sum_{k=1}^i {n \choose i}{i \choose k}\mathop{g_{\mu\nu}}\limits^{(n-i)}\mathop{g^{\alpha\beta}}\limits^{(i-k)}\\
&\cdot\biggr[\nabla_\alpha\nabla_\beta \mathop{f'}\limits^{(k)}-\sum_{l=1}^k {k \choose l}\mathop{{C_{\alpha\beta}}^\delta}\limits^{(k-l)}\nabla_\delta\mathop{f'}\limits^{(l)}\biggr]\\
&+\mathop{f'}\limits^{(0)}\mathop{R_{\mu\nu}}\limits^{(n)}-\frac{1}{2}\mathop{g_{\mu\nu}}\limits^{(n)}\mathop{f}\limits^{(0)}+\mathop{{C_{\mu\nu}}^\alpha}\limits^{(n)}\nabla_\alpha\mathop{f'}\limits^{(0)}\\
&+\sum_{i=1}^n {n \choose i}\mathop{g_{\mu\nu}}\limits^{(n-i)}\mathop{g^{\alpha\beta}}\limits^{(i)}\nabla_\alpha\nabla_\beta \mathop{f'}\limits^{(0)}\\
&-\sum_{i=1}^n\sum_{k=1}^i {n \choose i}{i \choose k}\mathop{g_{\mu\nu}}\limits^{(n-i)}\mathop{g^{\alpha\beta}}\limits^{(i-k)}\mathop{{C_{\alpha\beta}}^\delta}\limits^{(k)}\nabla_\alpha\mathop{f'}\limits^{(0)}
\end{eqnarray*}
Now, we replace (\ref{fn}) and (\ref{f'n}) in the term $n+1$
\begin{eqnarray*}
 \mathop{\Sigma_{\mu\nu}}\limits^{(n+1)}=&\sum_{i=1}^{n+1}\biggl[{n+1 \choose i}2i\mathop{R}\limits^{(i-1)}\mathop{R_{\mu\nu}}\limits^{(n-i+1)}\\
&-\frac{1}{2}{n+1 \choose i}\biggr(\mathop{R}\limits^{(i)}+i\sum_{m=0}^{i} {i-1 \choose m}\mathop{R}\limits^{(m)}\mathop{R}\limits^{(i-m-1)}\biggr)\mathop{g_{\mu\nu}}\limits^{(n-i+1)}\biggr]\\
&-2(n+1)\nabla_\mu\nabla_\nu \mathop{R}\limits^{(n)}+\sum_{i=1}^{n+1}{n+1 \choose i}\mathop{{C_{\mu\nu}}^\alpha}\limits^{(n-i+1)}\nabla_\alpha(2i\mathop{R}\limits^{(i-1)})\\
&+\sum_{i=1+1}^n\sum_{k=1}^i {n+1 \choose i}{i \choose k}\mathop{g_{\mu\nu}}\limits^{(n-i+1)}\mathop{g^{\alpha\beta}}\limits^{(i-k)}\\
&\cdot\biggr[\nabla_\alpha\nabla_\beta 2k\mathop{R}\limits^{(k-1)}-\sum_{l=1}^k {k \choose l}\mathop{{C_{\alpha\beta}}^\delta}\limits^{(k-l)}\nabla_\delta(2l\mathop{R}\limits^{(l-1)})\biggr]\\
&+\mathop{R_{\mu\nu}}\limits^{(n+1)}-\frac{1}{2}\mathop{g_{\mu\nu}}\limits^{(n+1)}\mathop{R}\limits^{(0)}.
\end{eqnarray*}
Now, we introduce the induction hypothesis, i.e. we use $\mathop{R_{\mu\nu}}\limits^{(i)}=0$ and $\mathop{R}\limits^{(i)}=0$ for all $i=0,\dots,n$. We obtain
\begin{equation}
\mathop{\Sigma_{\mu\nu}}\limits^{(n+1)}=\mathop{R_{\mu\nu}}\limits^{(n+1)}-\frac{1}{2}\mathop{g_{\mu\nu}}\limits^{(0)}\mathop{R}\limits^{(n+1)}=\mathop{G_{\mu\nu}}\limits^{(n+1)}.
\end{equation}
If we take the trace again, $\mathop{R_{\mu\nu}}\limits^{(n+1)}=0$ and $\mathop{R}\limits^{(n+1)}=0$.
Therefore, we have demonstrated that $\mathop{\Sigma_{\mu\nu}}\limits^{(n)}=\mathop{G_{\mu\nu}}\limits^{(n)}$ and, therefore,
\begin{equation}
 \bar{\Sigma}_{\mu\nu}=\mathop{\Sigma_{\mu\nu}}\limits^{(0)}+\lambda\mathop{\Sigma_{\mu\nu}}\limits^{(1)}+\frac{\lambda^2}{2!}\mathop{\Sigma_{\mu\nu}}\limits^{(2)}+\cdots=\mathop{G_{\mu\nu}}\limits^{(0)}+\lambda\mathop{G_{\mu\nu}}\limits^{(1)}+\frac{\lambda^2}{2!}\mathop{G_{\mu\nu}}\limits^{(2)}+\cdots=\bar{G}_{\mu\nu}
\end{equation}
Finally, we have obtained that the asymptotic expansions of these two tensors coincide, which indicates that they differ at most in singular terms on $\lambda$, which will be discussed below.
$\blacksquare$\\

\subsection{Case $f(R)=R+\lambda\Psi_\lambda(R)$}

Now, we want to generalize the previous results to the case where $f(R)=R+\lambda\Psi_\lambda(R)$, with $\Psi_\lambda$ an analytic function in a vicinity of $\lambda=0$. For such purpose, we expand this function around $\lambda=0$,
\begin{equation}
\bar{\Psi}=\mathop{\Psi}\limits^{(0)}+\lambda\mathop{\Psi}\limits^{(1)}+\frac{\lambda^2}{2!}\mathop{\Psi}\limits^{(2)}+\cdots,
\end{equation}
thus
\begin{eqnarray}\label{f0general}
          \mathop{f}\limits^{(0)}&=\mathop{R}\limits^{(0)}  \mbox{  for  } n=0,\\\label{fngeneral}
          \mathop{f}\limits^{(n)}&=\mathop{R}\limits^{(n)}+n\mathop{\Psi}\limits^{(n-1)}  \mbox{ for  } n\geq 1,
\end{eqnarray}
also
\begin{eqnarray}\label{f'0genral}
          \mathop{f'}\limits^{(0)}&=1  \mbox{  for  } n=0,\\ \label{f'ngenral}
          \mathop{f'}\limits^{(n)}&=n\mathop{\Psi'}\limits^{(n-1)}  \mbox{ for  } n\geq 1.
\end{eqnarray}
Now
\begin{eqnarray*}
 \Psi(\bar{R})=&\Psi(\mathop{R}\limits^{(0)}+\lambda\mathop{R}\limits^{(1)}+\frac{\lambda^2}{2!}\mathop{R}\limits^{(2)}+\cdots),\\
=&\Psi(\mathop{R}\limits^{(0)})+\lambda\frac{\partial \Psi}{\partial\bar{R}}\mathop{R}\limits^{(1)}+\frac{\lambda^2}{2!}\biggl[\frac{\partial^2 \Psi}{\partial\bar{R}^2}\mathop{R^2}\limits^{(1)}+\frac{\partial \Psi}{\partial\bar{R}}\mathop{R}\limits^{(2)}\biggr]\\
&+\frac{\lambda^3}{3!}\biggl[\frac{\partial^3 \Psi}{\partial\bar{R}^3}\mathop{R^3}\limits^{(1)}+3\frac{\partial^2 \Psi}{\partial\bar{R}^2}\mathop{R}\limits^{(2)}\mathop{R}\limits^{(1)}+\frac{\partial \Psi}{\partial\bar{R}}\mathop{R}\limits^{(3)}\biggr]+\cdots,
\end{eqnarray*}
therefore
\begin{equation}\label{eq:fff}
\eqalign{\mathop{f}\limits^{(0)}=\mathop{R}\limits^{(0)} \qquad &\mathop{f'}\limits^{(0)}=1,\\
    \mathop{f}\limits^{(1)}=\mathop{R}\limits^{(1)}+\Psi(\mathop{R}\limits^{(0)}) \qquad &\mathop{f'}\limits^{(1)}=\Psi'(\mathop{R}\limits^{(0)}),\\
    \mathop{f}\limits^{(2)}=\mathop{R}\limits^{(2)}+\Psi'(\mathop{R}\limits^{(0)})\mathop{R}\limits^{(1)} \qquad &\mathop{f'}\limits^{(2)}=2\Psi''(\mathop{R}\limits^{(0)})\mathop{R}\limits^{(1)}\\
    \vdots\qquad&\vdots.}
\end{equation}

The $n$-th term $\mathop{f}\limits^{(n)}$ is $\mathop{R}\limits^{(n)}$ plus a combinations of terms with derivatives of $\Psi$ with respect to $R$ and different orders of the Ricci scalar $\mathop{R}\limits^{(i)}$ for $i=0,\dots,n-1$. In the same way, the $n$-th term of $\mathop{f'}\limits^{(n)}$ is a combination of products with the Ricci scalar in different orders $\mathop{R}\limits^{(i)}$ for $i=0,\dots,n-1$.

\begin{theorem}
Let $\bar{\Sigma}_{ab}=0$ be the $f(R)$ MTG field equations in vacuum for the model $f(\bar{R})=\bar{R}+\lambda \Psi(\bar{R})$, where $\Psi(R)$ is analytic in a vicinity of $\lambda=0$, and $\Psi(0)=0$. Then, $\bar{\Sigma}_{ab}=\bar{G}_{ab}$ in vacuum.
\end{theorem}
Proof: Again we proceed by induction. We have at zero order
\begin{equation}
 \mathop{\Sigma_{\mu\nu}}\limits^{(0)}=\mathop{R_{\mu\nu}}\limits^{(0)}-\frac{1}{2}\mathop{g_{\mu\nu}}\limits^{(0)}\mathop{R}\limits^{(0)}=\mathop{G_{\mu\nu}}\limits^{(0)},
\end{equation}
whereas at first order,
\begin{equation}
 \mathop{\Sigma_{\mu\nu}}\limits^{(1)}=\mathop{R_{\mu\nu}}\limits^{(1)}-\frac{1}{2}\mathop{g_{\mu\nu}}\limits^{(0)}(\mathop{R}\limits^{(1)}+\mathop{\Psi}\limits^{(0)})-\nabla_\mu\nabla_\nu\mathop{\Psi'}\limits^{(0)} +\mathop{g_{\mu\nu}}\limits^{(0)}\square\mathop{\Psi'}\limits^{(0)}
\end{equation}
If $\Psi(0)=0$, we have
\begin{equation}
 \mathop{\Sigma_{\mu\nu}}\limits^{(1)}=\mathop{R_{\mu\nu}}\limits^{(1)}-\frac{1}{2}\mathop{g_{\mu\nu}}\limits^{(0)}\mathop{R}\limits^{(1)}=\mathop{G_{\mu\nu}}\limits^{(1)}
\end{equation}
Now, we suppose that the $n$-th order case is satisfied. Taking the trace in the equations $\mathop{G_{\mu\nu}}\limits^{(i)}=0$ we have $\mathop{R_{\mu\nu}}\limits^{(i)}=0$ and $\mathop{R}\limits^{(i)}=0$ for all $i=0,\dots,n$. Finally, using the properties (\ref{eq:fff}) we have,
\begin{equation}
\eqalign{\mathop{f}\limits^{(0)}=0 \qquad &\mathop{f'}\limits^{(0)}=1,\\
    \mathop{f}\limits^{(i)}=\mathop{R}\limits^{(i)} \qquad &\mathop{f'}\limits^{(i)}=0,}
\end{equation}
for $i=1,\dots,n$. In the case $n+1$, we have, from formula (\ref{ecufrn}),
\begin{eqnarray*}
 \mathop{\Sigma_{\mu\nu}}\limits^{(n+1)}=&\biggl[{n+1 \choose 0}\mathop{f'}\limits^{(0)}\mathop{R_{\mu\nu}}\limits^{(n+1)}-\frac{1}{2}{n+1 \choose n+1}\mathop{f}\limits^{(n+1)}\mathop{g_{\mu\nu}}\limits^{(0)}\biggr]\\
 =&\mathop{R_{\mu\nu}}\limits^{(n+1)}-\frac{1}{2}\mathop{R}\limits^{(n+1)}\mathop{g_{\mu\nu}}\limits^{(0)}=\mathop{G_{\mu\nu}}\limits^{(n+1)}.
\end{eqnarray*}
Thus $\mathop{\Sigma_{\mu\nu}}\limits^{(n)}=\mathop{G_{\mu\nu}}\limits^{(n)}$ and also $\mathop{R_{\mu\nu}}\limits^{(n+1)}=0$ and $\mathop{\Sigma_{\mu\nu}}\limits^{(n+1)}=0$. Finally
\begin{equation}
 \bar{\Sigma}_{\mu\nu}=\mathop{\Sigma_{\mu\nu}}\limits^{(0)}+\lambda\mathop{\Sigma_{\mu\nu}}\limits^{(1)}+\frac{\lambda^2}{2!}\mathop{\Sigma_{\mu\nu}}\limits^{(2)}+\cdots=\mathop{G_{\mu\nu}}\limits^{(0)}+\lambda\mathop{G_{\mu\nu}}\limits^{(1)}+\frac{\lambda^2}{2!}\mathop{G_{\mu\nu}}\limits^{(2)}+\cdots=\bar{G}_{\mu\nu}
\end{equation}
Therefore, we have found again that the asymptotic expansions for $\bar{\Sigma}_{\mu\nu}$ and $\bar{G}_{\mu\nu}$ are the same, which indicates that these two tensors differ at most in singular terms on $\lambda$. Furthermore, in the context of perturbative constraints, these results indicate that the approximations for the vacuum $f(R)$ field equations are equivalent to the approximations that could be done in GR, so, within this perturbative scheme, the MTG will not introduce any modification to the GR findings.

$\blacksquare$\\
Regarding the gauge freedom of perturbation theory, mentioned above, it is important to be aware that results such as the theorem just proved could be gauge-dependent in general. However, in this case, we are comparing theories in the vacuum, that is, $\Sigma(\lambda)_{\mu\nu}=0$ in each slide labelled by $\lambda$. This assumption is very important for our results since it allows us to disregard the gauge freedom problem by virtue of the generalized Stewart-Walker lemma. This lemma implies, for a $\Sigma_{\mu\nu}$ which vanishes in the background, that for any two different gauges $X$ and $Y$, $\mathop{\Sigma_X}\limits^{(n)}=\mathop{\Sigma_Y}\limits^{(n)}$ for any order $n$. The case of a non-vanishing energy momentum tensor $T_{\mu\nu}\neq0$ involves the construction of gauge-invariant quantities which are non-trivial and, therefore, it must be studied separately \cite{molano2020}. Nevertheless, in the following section, we discuss some remarks regarding this issue and its relation with the results here presented.

\section{Discussion}
In this work, we have constructed a perturbation theory for $f(R)$ gravity, applying the principles of perturbative constraints and following the guidelines proposed by Bruni and coworkers \cite{bruni:1997}. Although they already established that their formalism is equally applicable to any relativistic theory of gravity, we implemented their proposal explicitly for $f(R)$ models and supposed from the beginning that the comparison is to be made with GR. The introduced formalism can provide a foundation to approach further problems in relativistic perturbation theory such as the construction of such a theory in terms of gauge-invariant quantities. It is also relevant to note that our proposal is an alternative to the framework introduced by Capozziello and coworkers \cite{Capozziello_2008} with some advantages: we have obtained corrections at higher orders and, more importantly, we do not require spherically symmetric spacetimes, therefore we can reach a larger domain of applicability for the formalism. From this point of view, interesting results obtained within the framework of \cite{Capozziello_2008} can be reexamined with our formalism; as a example we can consider \cite{capozziello2012scalar}, where the relation of $f(R)$ gravity with scalar-tensor theories at the perturbative level is studied, as well as the consequences regarding the Birkhoff-Jebsen theorem.

Now, having discussed some features of our perturbational scheme, we should make some remarks with respect to the main result of this work; namely, the equivalence of $f(R)$ and GR perturbation theories in vacuum. Let us restrict ourselves to the spherically symmetric case reviewed before, in which the GR solution is the Schwarzschild metric. Our theorem implies, in this context, that $f(R)$ perturbation theory in vacuum will recover this metric as a solution. The fact that the Schwarzschild metric is a solution for many $f(R)$ models has been known in the literature through different methods, and it has even been argued that the Schwarzschild metric is the only black hole solution for a large class of $f(R)$ functions \cite{canate2016spherically}. One interesting consequence of our results within this framework is that we expect no effect on PPN parameters in the exterior of spherical masses in vacuum due to $f(R)$ corrections, which is interesting from an observational point of view.
In this context, it is remarkable to note that there are spherically symmetric solutions apart from the Schwarzschild solution in different $f(R)$ models. As an example, we recall \cite{cikintouglu2018vacuum}, where the Starobinsky model, $f(R)=R+\lambda R^2$, is considered and spherical symmetry is introduced as described above to study the exterior of a compact object. In this case we have that the field equations take the form
\begin{eqnarray}\label{ec:701}
 &2\lambda r\ddot{R}-(\dot{\eta}+\dot{\alpha})(1+2\lambda R+\lambda r\dot{R})=0,\\\label{ec:oscilador}
 &\lambda\ddot{R}+\lambda\dot{R}\biggl[\frac{1}{2}(\dot{\eta}-\dot{\alpha})+\frac{2}{r}\biggr]-\frac{Re^{\alpha}}{6}=0,\\\label{ec:ricciespherical}
 &R=-\frac{1}{2}\frac{2r^2\ddot{\eta}+r^2\dot{\eta}^2-r^2\dot{\eta}\dot{\alpha}+4\dot{\eta}r-4\dot{\alpha}r-4e^{\alpha}+4}{r^2e^{\alpha}},
\end{eqnarray}
where Eq. (\ref{ec:701}) is obtained from (\ref{ec:70}), Eq. (\ref{ec:oscilador}) is obtained from the trace equation (\ref{trace}), and Eq. (\ref{ec:ricciespherical}) is the expression for the Ricci scalar in terms of the metric functions.

Equation (\ref{ec:oscilador}), when divided by $\lambda$, resembles a harmonic oscillator equation
\begin{equation}\label{oscillatory2}
\ddot{R}+\dot{R}\biggl[\frac{1}{2}(\dot{\eta}-\dot{\alpha})+\frac{2}{r}\biggr]-\frac{Re^{\alpha}}{6\lambda}=0
\end{equation}
If $\lambda$ is negative and tends to zero, for some appropriate initial conditions, we expect oscillatory solutions with a high frequency. However, a negative $\lambda$ violates the local stability criterion \cite{zbMATH05782036}. Now, when $\lambda$ is positive and tends to zero, we do not have oscillatory solutions, but solutions involving increasing/decreasing exponentials. These solutions are not expected to reduce to the Schwarzschild solution when $\lambda$ tends to zero. In fact, if we return to Eqs. (\ref{ec:701}) and (\ref{ec:oscilador}), and let $\lambda$ to tend to zero, then we recover equations that lead to the Schwarzschild solution; namely, $\dot{\eta}+\dot{\alpha}=0$ and $R=0$. Thus, all the information with respect to the derivatives of $R$ is washed out. However, the boundary conditions regarding $R$ and its first derivative are still present in the problem and it is expected that the $\lambda=0$ solutions would violate such boundary conditions in many cases; this phenomenon is known as a \emph{boundary layer} for the perturbative problem. In this concrete example \cite{cikintouglu2018vacuum}, solutions are obtained, through the method of matched asymptotic expansions \cite{holmes2012introduction}, for spherically symmetric spacetimes outside a relativistic star; the solution for $R$ does not vanish, and includes decreasing exponential factors of the form $H\exp[-\sqrt{C/6\lambda}(r-R_s)]$, where $C$ and $H$ are integration constants and $R_s$ is the radius of the star. This form is to be expected from (\ref{ec:ricciespherical}), as discussed before. When analyzing the limit $\lambda\rightarrow0$ of this solution, it is found that the exponential factor tends to zero and this suppresses any behavior of the solution, recovering the $R=0$ behavior of GR \emph{except} at the point $r=R_s$, where boundary conditions are imposed for the solution which must be respected even for $\lambda\rightarrow0$; this implies that the limit spacetime of $f(R)$ solutions is not a solution of GR. One possibility to remedy this issue is to impose boundary conditions that resemble GR behavior; namely, $R=0$ on the boundary at $r=R_s$ in this case, but obviously this supposition excludes a large space of solutions which are acceptable from the point of view of $f(R)$ MTG; furthermore, it demands to involve the GR metric when defining the problem of finding solutions for $f(R)$ gravity, which is awkward from a mathematical point of view.

The described behavior indicates that there are solutions for $f(R)$ MTG field equations that are disconnected from their GR counterparts, in the sense that they can not be linked perturbatively with the latter in the framework of perturbative constraints. In other terms, our theorem implies that if $g_0$ is a solution of the Einstein field equations in the vacuum, then it is also a solution for $f(R)$ modified gravity field equations in $f(R)=R+\lambda \Psi(R)$ in vacuum; however, this is not the unique solution of the system, there can be another solution in this model which may not have as limit $g_0$ when $\lambda$ tends to zero. This implies that solutions different to GR solutions are disconnected from these solutions through $\lambda$ in these cases.

\begin{figure}[H]
\begin{center}
\includegraphics[width=0.6\linewidth]{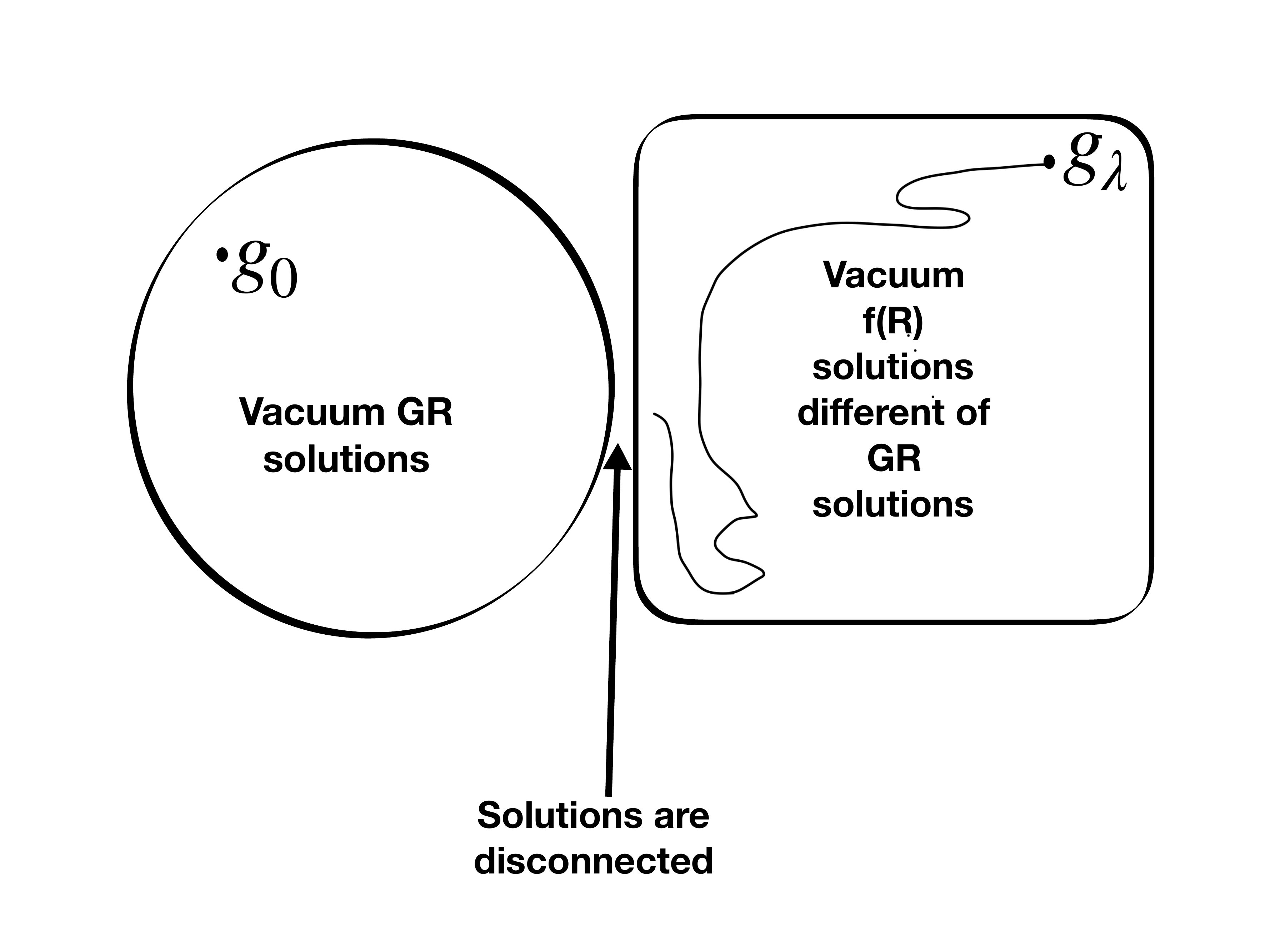}
\end{center}
\caption{Sketch of the solutions in $f(R)=R+\lambda \Psi(R)$ models of gravity. Any other solution different of GR in vacuum in these models are disconnected through $\lambda$}
\label{fig:sketch}
\end{figure}

As an additional example we can note that there are models where the Schwarzschild metric is not a solution but the corresponding solution is connected to the GR one, such as in the case $f(R)=R^{1+\lambda}$. The corresponding solution was provided by Clifton and Barrow \cite{PhysRevD.72.103005}. Here the solution tends to the Schwarzschild solution when $\lambda$ goes to zero. Our theorem can not be applied in this context since the function in the action is not of the form $f(R)=R+\lambda \Psi(R)$ clearly; thus, we can not use our perturbative technique in this kind of model. However, it is important to note that additional solutions to the Clifton and Barrow one are to be expected by virtue of the freedom in the additional boundary conditions of higher-derivative theories and these solutions could be non-connected. To show study this case, another technique must be used in addition to the described ones.

It could be argued that any $f(R)$ solution, disconnected from the corresponding GR one in the sense of perturbative constraints, should be discarded since it does not reproduce observations that support the latter; furthermore, if all the solutions for an specific $f(R)$ model fail to be connected, one would be tempted to believe that such model is not acceptable as a description of gravity. However, this reasoning entails a strong requirement for acceptable solutions that could be too demanding. It is natural to accept any solution which agrees with data up to the uncertainty range of the measurements, and, in this way, even disconnected solutions can be accepted. We must remind that $f(R)$ theories of gravity have associated four-order partial differential equations and therefore, we need information of the first, second and third derivatives of the metric on the boundary (or the metric and the connection in the Palatini approach), and this information implies the presence of new degrees of freedom that may disrupt the perturbative connection of the $f(R)$ solution with the GR one.

It is important to note that the results previously presented can be understood as a lack of effects beyond GR at any order at perturbation theory. In fact, the structure of the inductive proof of Theorem 3 indicates that there is some sort of deferment of the new effects to higher orders in perturbation theory. Concretely, at any order we have that the new effects are multiplied by $\lambda$, so they correspond to the following order. An essential feature to obtain this behavior is that the parameter accompanying the correction $\Psi(R)$ is the same parameter that defines the perturbation expansion, which is an essential feature of the perturbative constraints approach. There are explicit examples within this framework \cite{dedeo2008stable,cooney2009gravity,cooney2010neutron,ky2018perturbative,ky2019testing} that show that the absence of effects beyond GR does not occur in presence of matter; specifically, solutions have a non-trivial contributions from $\lambda$ terms, even though there are no new degrees of freedom in addition to the GR ones. An instructive example of the differences between the vacuum and non-vacuum cases is given in \cite{simon1992no}, where the cosmological inflation implemented with the Starobinsky model is studied in the context of perturbative constraints. In this work, it is found that no inflation can proceed in the absence of matter, excluding the possibility of a modified gravity source for the accelerated expansion within this model. In addition, it is shown that the presence of matter, in terms of a dark energy fluid, induces a new form for the metric which depends on the perturbation parameter $\lambda$ in a non-trivial way. Another interesting example is \cite{ky2018perturbative}, where explicit corrections are obtained for spherically symmetric metrics at first order for a variety of $f(R)$ functions, considering that there is matter present in the system. Again, considering vacuum and that the functions satisfy our conditions (namely $\Psi(0)=0$ in our notation), their results reproduce what is expected from our theorems, and show that these do not hold in the presence of matter. In conclusion, all these  results can be regarded as concrete realizations of our theorems with respect to the vacuum case, as well as counterexamples for any attempt to extend our theorem to the non-vacuum case with the same conditions stated above.

Any approach devoted to extend our results to the matter case will have to include additional restrictions on the possible theorems, in addition to a proper discussion of the possible gauge-invariant quantities that are relevant for that case. The corresponding formulation of a perturbation theory in terms of gauge-invariant quantities has been addressed by us in an upcoming paper \cite{molano2020}, as mentioned before. To summarize its results and their relation with the findings of the present work, let us remark that the construction of the gauge-invariant variables follows closely the corresponding process of GR (see \cite{nakamura2019second} for a recent review), and the main differences arise from the structure of the $f(R)$ field equations, which is clearly more complicated. The resulting formalism is applied to cosmology, obtaining equations that could be used to study problems such as large-scale structure and cosmological magnetic fields. For the time being, let us emphasize that such formalism constitutes an excessive complication for the scope of this paper, since the generalized Stewart-Walker lemma guarantees that the perturbations involved in our theorems are gauge-invariant by themselves. In addition, the matter case, where the construction of gauge-invariant quantities is more complex, is not relevant in this context since we already know that our theorems are not satisfied in such case, as argued in the previous paragraph.

In a related sense with the previous remarks, we can provide some insight into the presence of non-trivial effects beyond GR by studying the implications of the negation of Theorem 3. Namely, we can suppose that there are differences between the perturbation theories of some $f(R)$ model and GR; therefore, one of the following possibilities must be true:
\begin{enumerate}
    \item There is matter in the case under study.
    \item The perturbative expansions used have a different form from the one used in perturbative constraints, which is analytic on $\lambda$.
    \item The function $f(R)$ is not of the form $R+\lambda \Psi(R)$.
\end{enumerate}
In addition to the situation discussed above, the first possibility can be understood as that the coupling of matter with gravity captures the additional structure introduced by the $f(R)$ model, even as no new gravitational degree of freedom appears; in fact, let us remark that the perturbative constraints approach eliminates the additional degree of freedom expected from the $f(R)$ field equations, regardless of the presence of matter. Similar results are known in quantum theory of higher-derivative theories \cite{cheng2002perturbative}. With respect to the second possibility, we must note that this is the relevant case for a problem with boundary layers such as \cite{cikintouglu2018vacuum}. Actually, the matched asymptotic expansions method takes into account the boundary conditions by employing expansions on the so-called fast variables, which diverge as $\lambda$ tends to zero. This is the reason behind that solutions different from Schwarzschild can be obtained even in a perturbative setting. Other alternatives in this context would be that the parameter that multiplies the additional term in the gravitational action is not the same parameter that defines the perturbative series; as examples of this approach we can count \cite{berry2011linearized}, in the context of gravitational waves, and \cite{suvorov2019gravitational}, which studies the perturbations of Kerr black holes. Finally, our third possibility can be important for cases like Clifton and Barrow solution \cite{PhysRevD.72.103005}, where the function $f(R)$ is not of the form required by our theorem, as remarked before.

Finally, an important feature of our results that must be noted is their generality with respect to symmetries. Therefore we can apply it to problems beyond spherical symmetry, such as spacetimes with gravitational waves. We leave for future work the study of the implications of our method and results for this setting, as well as the problem of the construction of gauge invariant quantities in this context.

\section{Conclusions}
In this work, we constructed a method to compare solutions in modified theories of gravity with General Relativity solutions, and with it, we implemented a perturbational scheme, based on the perturbative constraints method, to construct the solutions in such theories as corrections to General Relativity results; such scheme generalizes previous research and opens up the possibility of studying issues such as the gauge invariance of the calculations. We used the proposed formalism in the case of $f(R)$ modified gravity; in particular, for the model $f(R)=R+\lambda \Psi(R)$ in vacuum. Within our assumptions, we proved that vacuum perturbation theory in the considered $f(R)$ models is completely equivalent to General Relativity perturbation theory, therefore perturbative corrections to solutions to the latter are not to be expected within the conditions of our theorem. The implications of this result and its connection with previous research in the literature were discussed.
\section*{Acknowledgements}
The authors are indebted to Luz \'Angela Garc\'ia for her careful revision of the manuscript and valuable suggestions. F. D. V. acknowledges funding from the School of Sciences of Universidad de los Andes during the realization of this work, and also acknowledges the hospitality received during his stay at Observatorio Astron\'omico Nacional of Universidad Nacional de Colombia. P. B. is funded by the Beatriz Galindo contract BEAGAL 18/00207 (Spain).

\section*{References}

\bibliography{Bibliography}

\providecommand{\newblock}{}
\begin{thebibliography}{10}
\expandafter\ifx\csname url\endcsname\relax
  \def\url#1{{\tt #1}}\fi
\expandafter\ifx\csname urlprefix\endcsname\relax\def\urlprefix{URL }\fi
\providecommand{\eprint}[2][]{\url{#2}}

\bibitem{sotiriou2010f}
Sotiriou T~P and Faraoni V 2010 {\em Reviews of Modern Physics\/} {\bf 82} 451

\bibitem{nojiri2011unified}
Nojiri S and Odintsov S~D 2011 {\em Physics Reports\/} {\bf 505} 59--144

\bibitem{nojiri2017modified}
Nojiri S, Odintsov S and Oikonomou V 2017 {\em Physics Reports\/} {\bf 692}
  1--104

\bibitem{woodard2007avoiding}
Woodard R 2007 Avoiding dark energy with 1/${R}$ modifications of gravity {\em
  The Invisible Universe: Dark Matter and Dark Energy\/} (Springer) pp 403--433

\bibitem{PhysRevD.72.103005}
Clifton T and Barrow J~D 2005 {\em Phys. Rev. D\/} {\bf 72}(10) 103005
  \urlprefix\url{https://link.aps.org/doi/10.1103/PhysRevD.72.103005}

\bibitem{Capozziello_2008}
Capozziello S, Stabile A and Troisi A 2008 {\em Classical and Quantum
  Gravity\/} {\bf 25} 085004
  \urlprefix\url{https://doi.org/10.1088/0264-9381/25/8/085004}

\bibitem{Sebastiani_2011}
Sebastiani L and Zerbini S 2011 {\em The European Physical Journal C\/} {\bf
  71} ISSN 1434-6052
  \urlprefix\url{http://dx.doi.org/10.1140/epjc/s10052-011-1591-8}

\bibitem{simon1990higher}
Simon J~Z 1990 {\em Physical Review D\/} {\bf 41} 3720

\bibitem{jaen1986reduction}
Ja{\'e}n X, Llosa J and Molina A 1986 {\em Physical Review D\/} {\bf 34} 2302

\bibitem{eliezer1989problem}
Eliezer D and Woodard R 1989 {\em Nuclear Physics B\/} {\bf 325} 389--469

\bibitem{simon1991stability}
Simon J~Z 1991 {\em Physical Review D\/} {\bf 43} 3308

\bibitem{glavan2018perturbative}
Glavan D 2018 {\em Journal of High Energy Physics\/} {\bf 2018} 136

\bibitem{mottola2017scalar}
Mottola E 2017 {\em Journal of High Energy Physics\/} {\bf 2017} 43

\bibitem{simon1992no}
Simon J~Z 1992 {\em Physical Review D\/} {\bf 45} 1953

\bibitem{dedeo2008stable}
DeDeo S and Psaltis D 2008 {\em Physical Review D\/} {\bf 78} 064013

\bibitem{cooney2009gravity}
Cooney A, DeDeo S and Psaltis D 2009 {\em Physical Review D\/} {\bf 79} 044033

\bibitem{castellanos2018higher}
Castellanos A~R~R, Sobreira F, Shapiro I~L and Starobinsky A~A 2018 {\em
  Journal of Cosmology and Astroparticle Physics\/} {\bf 2018} 007

\bibitem{Solomon_2018}
Solomon A~R and Trodden M 2018 {\em Journal of Cosmology and Astroparticle
  Physics\/} {\bf 2018} 031--031
  \urlprefix\url{https://doi.org/10.1088\%2F1475-7516\%2F2018\%2F02\%2F031}

\bibitem{barros2019bouncing}
Barros B~J, Teixeira E~M and Vernieri D 2019 {\em arXiv preprint
  arXiv:1907.11732\/}

\bibitem{cooney2010neutron}
Cooney A, DeDeo S and Psaltis D 2010 {\em Physical Review D\/} {\bf 82} 064033

\bibitem{molano2020}
Molano D, Villalba F~D, Casta\~neda L and Bargue\~no P 2020 Gauge-invariant
  perturbation theory in $f({R})$ theories of gravity (in preparation)

\bibitem{bruni:1997}
Bruni M, Matarrese S, Mollerach S and Sonego S 1997 {\em Classical and Quantum
  Gravity\/} {\bf 14} 2585
  \urlprefix\url{http://iopscience.iop.org/0264-9381/14/9/014}

\bibitem{raey}
Guarnizo A, Casta\~neda L and Tejeiro J~M 2010 {\em General Relativity and
  Gravitation\/} {\bf 42} 2713--2728 ISSN 0001-7701
  \urlprefix\url{http://dx.doi.org/10.1007/s10714-010-1012-6}

\bibitem{padmanabhan2016atoms}
Padmanabhan T 2016 {\em International Journal of Modern Physics D\/} {\bf 25}
  1630020

\bibitem{wald:1984}
Wald R 1984 {\em General Relativity\/} (The University of Chicago Press)

\bibitem{Sebastiani2011}
Sebastiani L and Zerbini S 2011 {\em The European Physical Journal C\/} {\bf
  71} 1591 ISSN 1434-6052
  \urlprefix\url{https://doi.org/10.1140/epjc/s10052-011-1591-8}

\bibitem{zbMATH05782036}
{Capozziello} S and {Faraoni} V 2011 {\em {Beyond Einstein gravity. A survey of
  gravitational theories for cosmology and astrophysics.}\/} (New York, NY:
  Springer) ISBN 978-94-007-0164-9/hbk; 978-94-007-0165-6/ebook

\bibitem{Starobinsky198099}
Starobinsky A 1980 {\em Physics Letters B\/} {\bf 91} 99 -- 102 ISSN 0370-2693
  \urlprefix\url{http://www.sciencedirect.com/science/article/pii/037026938090670X}

\bibitem{cikintouglu2018vacuum}
{\c{C}}{\i}k{\i}nto{\u{g}}lu S 2018 {\em Physical Review D\/} {\bf 97} 044040

\bibitem{geroch:1969}
Geroch R 1969 {\em Communications in Mathematical Physics\/} {\bf 13} 180--193
  \urlprefix\url{http://projecteuclid.org/euclid.cmp/1103841574}

\bibitem{nakamura2019second}
Nakamura K 2019 {\em arXiv preprint arXiv:1912.12805\/}

\bibitem{sachs:1964}
Sachs R~K 1964 {\em Relativity Groups and Topology\/}

\bibitem{capozziello2012scalar}
Capozziello S and S{\'a}ez-G{\'o}mez D 2012 {\em Annalen der Physik\/} {\bf
  524} 279--285

\bibitem{canate2016spherically}
Ca{\~n}ate P, Jaime L~G and Salgado M 2016 {\em Classical and Quantum
  Gravity\/} {\bf 33} 155005

\bibitem{holmes2012introduction}
Holmes M~H 2012 {\em Introduction to perturbation methods\/} vol~20 (Springer
  Science \& Business Media)

\bibitem{ky2018perturbative}
Ky N~A, Van~Ky P and Van N~T~H 2018 {\em The European Physical Journal C\/}
  {\bf 78} 539

\bibitem{ky2019testing}
Ky N~A, Van~Ky P and Van N~T~H 2019 {\em arXiv preprint arXiv:1904.04013\/}

\bibitem{cheng2002perturbative}
Cheng T~C, Ho P~M and Yeh M~C 2002 {\em Nuclear Physics B\/} {\bf 625} 151--165

\bibitem{berry2011linearized}
Berry C~P and Gair J~R 2011 {\em Physical Review D\/} {\bf 83} 104022

\bibitem{suvorov2019gravitational}
Suvorov A~G 2019 {\em Physical Review D\/} {\bf 99} 124026

\end{thebibliography}


\end{document}